\documentclass[twocolumn]{aastex63}
\usepackage{amsmath}
\usepackage{threeparttable}
\usepackage{lineno}
\usepackage{cleveref}
\usepackage{xcolor, ulem}
\newcommand{\add}[1]{\textcolor{black}{#1}}

\newcommand{\mathendash}{\,\text{\textendash}\,}

\shorttitle{Can Ap/Bp Stars Explain the Orbital Decay of Algol-type Binaries?}
\shortauthors{Wu et al.}

\begin{document}

\title{Can the Anomalous Magnetic Braking of Ap/Bp Stars Explain the Orbital Decay of Algol-type Binaries?}

\author[0009-0009-5668-2147]{Guan-Long Wu}
\affil{Ministry of Education Key Laboratory for Nonequilibrium Synthesis and Modulation of Condensed Matter, School of Physics, Xi’an Jiaotong University, Xi’an 710049, People’s Republic of China; zuozyu@xjtu.edu.cn}
\author[0000-0001-6693-586X]{Zhao-Yu Zuo}
\affil{Ministry of Education Key Laboratory for Nonequilibrium Synthesis and Modulation of Condensed Matter, School of Physics, Xi’an Jiaotong University, Xi’an 710049, People’s Republic of China; zuozyu@xjtu.edu.cn}
\author[0000-0002-0785-5349]{Wen-Cong Chen}
\affil{School of Science, Qingdao University of Technology, Qingdao 266525, People’s Republic of China;chenwc@pku.edu.cn}

\begin{abstract}
\add{Several Algol-type binaries were detected to be experiencing rapid orbital decay, which is in contradiction with the conservative mass transfer model.}
In this work, we investigate whether anomalous magnetic braking (MB) of intermediate-mass Ap/Bp stars, characterized by surface magnetic fields of $\sim10^2 \mathendash 10^4~\rm G$, could drive orbital decay through magnetic wind-field coupling. Using the stellar evolution code {\ttfamily MESA}, we simulate the evolution of several \add{main-sequence binaries containing Ap/Bp stars}, with typical initial parameters \add{of Algol binaries}. Our models indicate that the anomalous MB mechanism could induce orbital decay in long timescales (hundreds of Myr to several Gyr), \add{reproducing several basic Algol parameters such as the effective temperatures and luminosities of donor stars.
However, the predicted orbital period decay rates are much lower than those observed in several Algol systems. 
We analyze the limitations of the anomalous MB model and discuss alternative mechanisms that could account for the long- or short-term orbital period variations observed in Algol systems, including a surrounding circumbinary disk, stellar expansion, the Applegate mechanism, and the light travel-time effect. Long-term observations are still required to distinguish between these mechanisms in the future.}
\end{abstract}


\section{Introduction}
Algol-type binaries are semi-detached binary systems with orbital periods spanning several hours to tens of days, where a Roche lobe-filling K/F-type giant/subgiant ($M_{\rm donor}$) transfers material to a hotter A/B-type main-sequence (MS) companion ($M_{\rm accretor}$) through Roche lobe overflow \citep[RLOF,][]{kopa55,giur83}. Notably, the initial mass of the donor star exceeds that of the A/B-type MS companion. Mass transfer begins via thermal-timescale-dominated RLOF ($\tau_{\rm th} \sim 10^5\mathendash 10^6~\rm yr$) when the material flows from the more massive to the less massive component. Subsequently, when the mass ratio ($q \equiv M_{\rm donor}/M_{\rm accretor}$) inverts below unity ($q < 1$), the system enters nuclear-timescale mass transfer ($\tau_{\rm nuc} \sim 10^7\mathendash 10^8~\rm yr$), marking its transition to the canonical Algol configuration \citep{morton1960evolutionary}.

Algol-type binaries are ideal astrophysical laboratories for probing mass transfer processes and angular momentum (AM) evolution. These systems critically test stellar and binary evolution models through wind-driven mass loss, nonconservative mass transfer, and AM redistribution mechanisms. Under conservative mass transfer, orbital expansion is theoretically expected as lower-mass donors transfer material to more massive accretors \citep{Huang1963}. \add{However, systemic mass/AM loss during binary evolution is evidenced by both multi-wavelength observations \citep{crawford1955subgiant,refs74,mass75} and the significantly lower orbital AM of semi-detached binaries compared to detached systems \citep{chau79}.}
By comparing the C abundance predicted by the theoretical models with that of the observational analysis, \cite{sarn93} further constrained that about $15\%$ of the initial total mass and $30\%$ of the initial AM of the Algol system Beta Per should be lost. Using a {$\chi^{2}$-minimizing} procedure, \cite{nels01} suggested that hot Algols follow a near-conservative evolution, while cool systems exhibit substantial AM loss.

Observationally, some Algol binaries have been found to undergo rapid orbital decay \citep{Qian2000a,Qian2000b,Qian2001a,Qian2001b,Qian2001c,Qian2002,LloydGuilbault2002, QianBoon2003}. However, conventional AM loss channels \textemdash{} direct mass loss and standard magnetic braking (MB) \textemdash{} are inadequate to account for the detected orbital decay in these systems \citep{Chen2006}. A surrounding circumbinary disk was proposed as an efficient mechanism for extracting AM, driving the orbital decay of Algol binaries \citep{Chen2006,Ibano2006}. 
\add{Furthermore, detailed analyses of the $O-C$ curves for some Algol binaries frequently reveal that long-term orbital period variations are superimposed with cyclic variations \citep{Wang2019} or exhibit purely cyclic variations \citep{Yang2014AFGem}. Two primary mechanisms have been proposed to account for these orbital period modulations: (1) the gravitational coupling of the binary orbit to the deformation of a magnetically active donor star \citep{appl92}, and (2) the light travel-time effect \citep[LTTE,][]{Irwin1952} caused by a third body. Apsidal motion \citep{Sterne1939Apsidal} is generally excluded as a viable explanation for such cyclic variations in Algol binaries, as tidal interactions and mass transfer processes are expected to circularize orbits \citep{Ma2022TWCas}.}

At present, the secular orbital decay of Algol binaries is not yet understood well, and other alternative AM loss mechanisms cannot be completely ruled out. Some previous studies have suggested that the anomalous MB of Ap/Bp stars with an anomalously strong magnetic field can efficiently extract orbital AM from binary systems, \add{thereby} driving a rapid orbital decay of X-ray binaries in an adequately long timescale \citep{Justham2006,Chen2016}. \add{This mechanism has been proposed as a possible explanation for the rapid orbital decay observed}
in black hole low-mass X-ray binaries such as XTE J1118+480, A0620-00 \citep{gonz14}, and Nova Muscae 1991 \citep{gonz17}. Recently, \cite{chen24} found that the anomalous MB of Ap/Bp stars could produce the rapid orbital decay observed in M82 X-2.

In this work, we attempt to diagnose whether the orbital decay observed in \add{four} Algol binaries could be interpreted by the anomalous MB of Ap/Bp stars. In section \ref{section:Input_Physics}, we describe the stellar evolution code and the input physics. The numerical results for several Algol systems are presented in Section \ref{section:Numerical_Results}. 
\add{In Section \ref{section:Discussion}, we analyse the limitations of the anomalous MB model and discuss alternative mechanisms. Finally, the summary is given in Section \ref{section:Summary}.}

\section{Stellar evolution code and anomalous MB model} \label{section:Input_Physics}
\subsection{Stellar evolution code}
\add{To study the influence of anomalous MB of Ap/Bp stars on binary evolution, and to examine whether it can account for the rapid orbital decay observed in several Algol binaries}, we employ the binary module in the Modules for Experiments in Stellar Astrophysics \citep[{\ttfamily MESA} version r24.08.1,][] {Paxton2011,Paxton2013,Paxton2015,Paxton2018,Paxton2019,Jermyn2023} to model the evolution of several zero-age MS binary systems \add{containing Ap/Bp stars}. The initial system consists of a more massive \add{Ap/Bp star} (star I, with a mass of $M_1$) and a less massive MS star (star II, with a mass of $M_2$) in a circular orbit. For simplicity, the {\ttfamily MESA} code only models the nuclear synthesis and evolution of star I with a solar composition (i.e., $X = 0.7, Y = 0.28, Z = 0.02$), and star II is considered as a point mass. 

Once star I fills the Roche lobe, the mass transfer rate ($\dot{M}_{\rm tr}$) is calculated using the "Kolb" mass-transfer scheme \citep{kolb90}. \add{The mass transfer process is assumed to be conservative, i.e., the accretion rate of the accretor is $\dot{M}_2=\dot{M}_{\rm tr}$.} The anomalous MB model requires a wind from star I, for which we take the ``Dutch'' wind scheme with a scaling factor of 1.0 including \texttt{hot\_wind\_scheme}, \texttt{cool\_wind\_RGB\_scheme}, and \texttt{cool\_wind\_AGB\_scheme} \citep{Glebbeek2009} during its evolution. The Type 2 opacities are used for extra C/O burning after star I initiates the He burning.

The loss of orbital AM plays a key role in influencing the evolution of Algol binaries. In this work, we consider three AM loss mechanisms as follows: anomalous MB, stellar winds, and gravitational radiation. It is clear that the effect of gravitational radiation is much weaker than those of the other two mechanisms. 

\subsection{Anomalous MB model}
In observations, the spin periods of some MS stars with convective envelopes were found to lengthen rapidly with age \citep{kraf67}. This phenomenon should arise from the coupling between the stellar winds and the magnetic field \citep{VerbuntZwaan1981}. The charged particles are thought to be tied to the magnetic field lines and kept in corotation with the star until the magnetospheric radius $r_{\rm m}$ ($\gtrsim 5\mathendash 10~\text{stellar radii}$), resulting in a significant loss of AM \citep{mest68,skum72}. With the spin-down of the evolved star, the tidal interaction between the two components in a close binary would continuously spin the star back up into synchronous rotation with the orbital motion. Therefore, the MB process indirectly consumes the orbital AM of the binary system. It was realized that in cataclysmic binaries and low-mass X-ray binaries, the MB mechanism could produce a mass transfer rate compatible with observations \citep{VerbuntZwaan1981,patt84}. 

The canonical MB mechanism cannot operate for stars with mass $\ga 1.5~M_\odot$, in which the radiative envelopes cannot produce a dynamo effect \citep{Kawaler1988}. However, \cite{Justham2006} proposed an anomalous MB scenario of Ap/Bp stars with an intermediate-mass ($\ga 1.5~M_\odot$) to account for the formation of black hole low-mass X-ray binaries. Those so-called Ap and Bp stars exhibit some observational evidence of anomalously strong surface magnetic fields of $\sim 10^2\mathendash 10^4~\rm G$, which may be a 'fossil' magnetic field, i.e., the remnants of the star formation \citep{moss89,brai04}. The Ap/Bp stars were also proposed to be the merger products of two (pre-) MS stars, and their magnetic fields are thought to originate from dynamo action during and shortly after the merger \citep{Ferrario2009,Schneider2016,Schneider2019}. 
{Among those MS stars with spectral ranges A, B, and O, about $7\mathendash 10\%$ host strong, large-scale magnetic fields \citep{Donati2009Magnetic,Ferrario2015Magnetic,Fossati2015}.} The spectral types of Ap/Bp stars are mainly concentrated in the ranges of A2\,\textendash\,B5, with masses and surface magnetic fields in the ranges of $1.6\mathendash 5~M_\odot$ and $500\mathendash 35,000~\rm G$ \citep{taur23}, respectively.

\add{In the anomalous MB model proposed by \citet{Justham2006}, stellar winds are driven from the donor star by X-ray irradiation originating from the accreting compact object.} However, the two components of Algol systems and their progenitors are MS stars. Therefore, we invoke intrinsic stellar winds from the evolved star instead of the irradiation-driven winds. Assuming that the stellar winds depart from the magnetic lines at the magnetospheric radius, they carry away the specific AM of the magnetospheric radius. As a consequence, the loss rate of orbital AM caused by the anomalous MB is given by
\begin{equation}
\dot{J}_{\rm MB}=-\dot{M}_{\rm w}r_{\rm m}^{2}\frac{2\pi}{P}, \label{eq:J_MB}
\end{equation}
where $\dot{M}_{\rm w}$ is the loss rate of stellar winds of the evolved star, $r_{\rm m}$ is the magnetospheric radius, and $P$ is the orbital period. As a location where the stellar magnetic energy density (magnetic pressure) is equal to the kinetic energy density (ram pressure) of the stellar winds, the magnetospheric radius can be expressed as
\begin{equation}
r_{\rm m}\simeq B_{\rm s}^{1/2}R_1^{13/8}\dot{M}_{\rm w}^{-1/4}\left( GM_1 \right)^{-1/8}, \label{eq:r_m}
\end{equation}
where $B_{\rm s}$, $M_1$, and $R_1$ are the surface magnetic field, the mass, and the radius of the evolved star, respectively. Inserting equation (\ref{eq:r_m}) into equation (\ref{eq:J_MB}), we can derive the loss rate of AM as
\begin{equation}
\dot{J}_{\rm MB}=-\frac{2\pi B_{\rm s}R_1^{13/4}\dot{M}_{\rm w}^{1/2}}{\left( GM_1 \right) ^{1/4}P}.
\end{equation}
For the same evolved star, the loss rate of orbital AM is related to the magnetic field, the loss rate of stellar winds, and the orbital period.
For different evolved stars, $\dot{J}_{\rm MB}$ strongly depends on the stellar radius, while it weakly depends on the stellar mass.

\subsection{Orbital evolution of MS binaries} \label{subsection: Wind}
The total orbital AM of a binary system with a circular orbit is $J=2\pi \mu a^2/P$, where $\mu=M_1M_2/(M_1+M_2)$ is the reduced mass and $a$ is the orbital separation. After the RLOF, the mass loss rate of the evolved star I includes the mass transfer rate ($\dot{M}_{\rm tr}$) and the loss rate ($\dot{M}_{\rm w}$) of stellar winds, i.e. $-\dot{M}_{1}=\dot{M}_{\rm tr}+\dot{M}_{\rm w}$. Assuming that the accreting efficiency of the accretor is $\beta=-\dot{M}_2/\dot{M}_1=-\dot{M}_{\rm tr}/\dot{M}_1$, the orbital period derivative obeys the following equation as
\begin{equation}
\frac{\dot{P}}{P}=3\frac{\dot{J}}{J}-3\frac{\dot{M}_1}{M_1}\left[1-q\beta-\frac{q(1-\beta)}{3(1+q)}\right]
    , \label{eq:pdot}
\end{equation}
where $q=M_1/M_2$ is the mass ratio of the binary. Ignoring the loss of stellar winds ($\dot{M}_{\rm w}=0$), we have $\beta=1$. When $q>1$, the second term on the right-hand side of the equation (\ref{eq:pdot}) causes an orbital decay effect, i.e., the mass transfer from the more massive evolved star to the less massive accretor tends to shorten the orbit. In contrast, the mass transfer would widen the orbit as $q<1$. As a critical point, it achieves a minimum period at $q=1$ if $\dot{J}=0$. In principle, a binary system should evolve to a minimum period at $q<1$ when the loss of AM due to the anomalous MB mechanism is included.

\begin{figure}
  \centering
  \includegraphics[width=1.0\columnwidth,trim={0 0 0 0},clip]{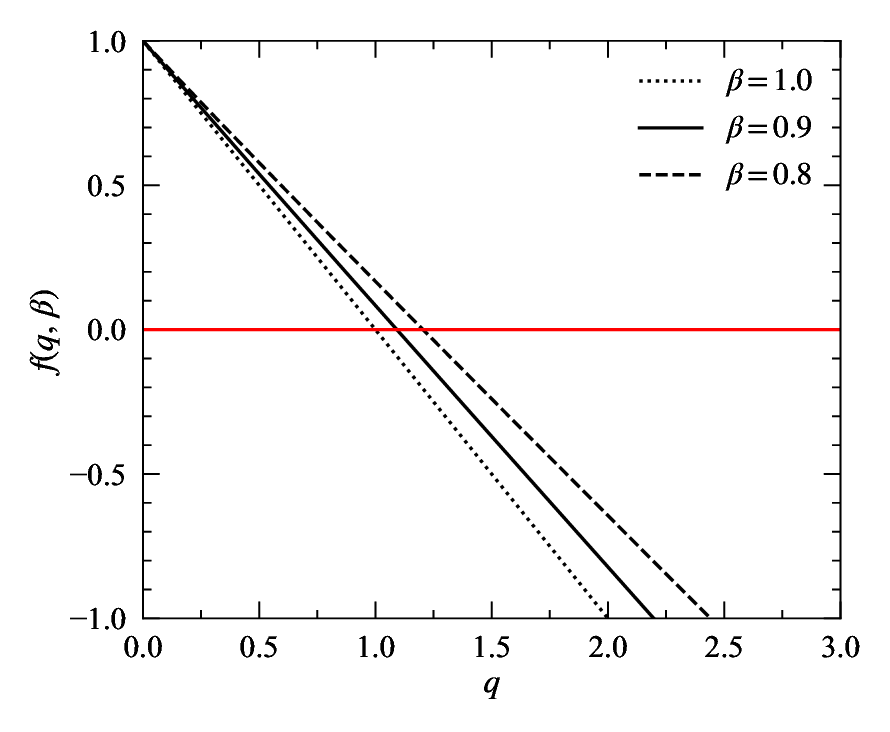}
  \caption{Relation between the function $f(q,\beta)$ and the mass ratio $q$. The dotted, solid, and dashed lines describe the cases with $\beta=1.0, 0.9$, and $0.8$, respectively. The horizontal red line denotes $f(q,\beta)=0$, over which the mass transfer would produce an orbital expansion effect.} \label{fig:fq}
\end{figure}

Actually, our detailed binary evolution models find that the first minimum period exhibits at $q>1$. This phenomenon is caused by the stellar winds of the evolved star. Taking $f(q,\beta)=1-q\beta-q(1-\beta)/3(1+q)$, in Figure \ref{fig:fq} we plot the evolution of $f(q,\beta)$ as a function of mass ratio $q$ for different values of accreting efficiency $\beta$. It is clear that $f(q,\beta)=0$ occurs at $q>1$ when $\beta=0.8$ and $0.9$ (i.e. $-\dot{M}_{\rm w}=0.2\dot{M}_1$ and $0.1\dot{M}_1$, respectively). Since $\dot{M}_1<0$, the mass transfer with a positive $f(q,\beta)$ naturally leads to an orbital expansion when $q>1$. As $\beta=1$, $f(q,\beta)=0$ occurs at $q=1$, hence the minimum period is achieved at $q=1$.

In general, the loss of AM (see also the first term on the right-hand side of Equation \ref{eq:pdot}) produces an orbital decay effect. However, a positive $f(q,\beta)$ results in a positive second term on the right-hand side of Equation (\ref{eq:pdot}), i.e., the mass transfer causes an orbital expansion effect. Therefore, the fates of the orbital evolution of Algol binaries depend on the ongoing competition between these two effects. 

\begin{figure*}[htb]
  \centering
  \includegraphics[width=1\textwidth,trim={0 0 0 0},clip]{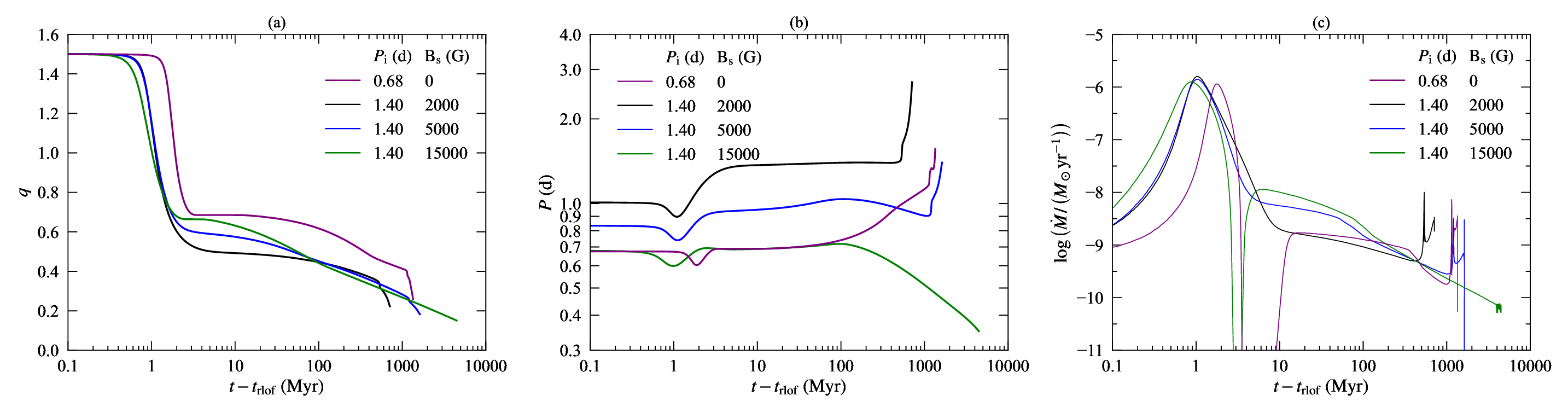}
  \caption{Evolution of the mass ratio (panel a), orbital period (panel b), and mass-transfer rate (panel c) as a function of the mass-transfer timescale ($t-t_{\rm rlof}$, $t_{\rm rlof}$ is the stellar age when the Roche lobe overflow occurs) for MS binary systems with $(M_1,M_2)=(3.0,2.0)~M_\odot$. The purple curve corresponds to the initial orbital period $P_{\rm i} = 0.68~\rm days$ under the standard MB law (i.e. $B_{\rm s}=0~\rm G$). The black, blue, and green curves correspond to $P_{\rm i} = 1.4~\rm days$ and $B_{\rm s}=2000~\rm G$, $5000~\rm G$, and $15000~\rm G$, respectively.} \label{fig:P_1_4}
\end{figure*}

\section{Simulated Results} \label{section:Numerical_Results}
\add{To systematically study the influence of anomalous MB mechanism on binary evolution, as well as its potential to account for the observed orbital decay in several Algol binaries, we conduct detailed binary evolution calculations for MS binaries with fixed initial masses of $(M_1, M_2) = (3.0, 2.0)~M_\odot$. We first study the influence of evolved stars' surface magnetic fields and initial orbital periods ($P_{\rm i}=1.4, 2.4$ and $3.4~\rm days$). Subsequently, we compare our simulated results with the observational data in the HR diagram and the $-\dot{P}$ versus mass transfer timescale diagram. Finally, we analyze why the anomalous MB mechanism fails to reproduce the orbital decay rates observed in several Algol binaries, from the perspective of evolved stars' stellar wind evolution.}

\begin{figure}
  \centering
  \includegraphics[width=1\columnwidth,trim={0 0 0 0},clip]{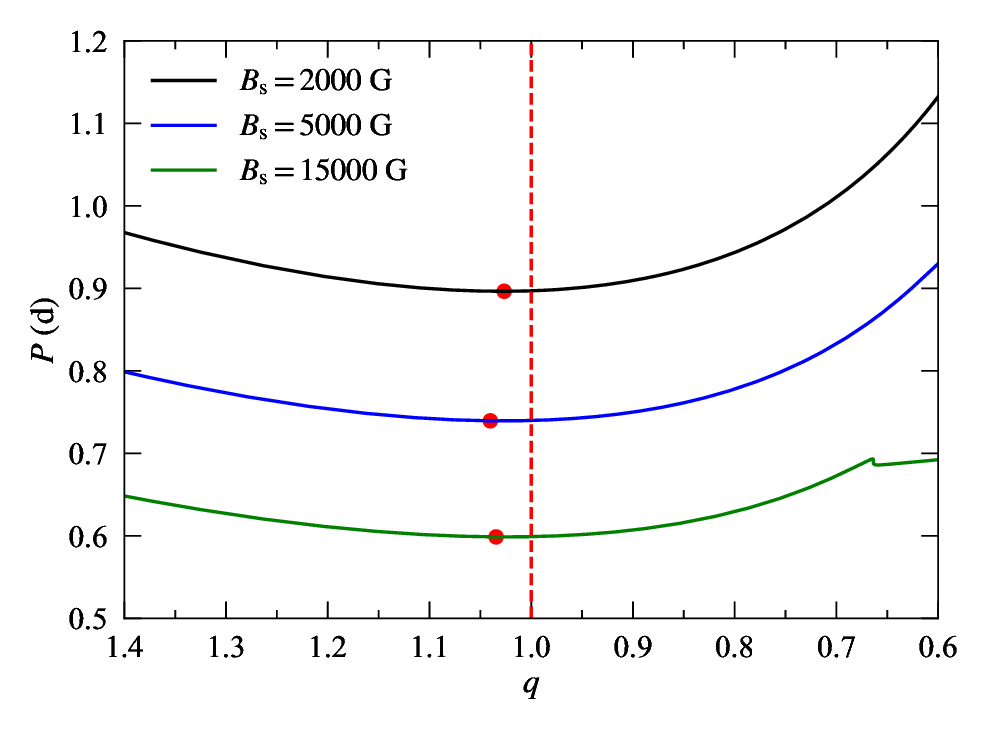}
  \caption{Evolution of the orbital period as a function of mass ratio for the same three systems with Ap/Bp components as in Figure \ref{fig:P_1_4}. Here, the standard MB law (i.e., $B_{\rm s}=0~\rm G$) was ignored. The red solid circles denote the first minimum orbital periods.} \label{fig:P_q}
\end{figure}

\subsection{Influence of surface magnetic fields}
Figure \ref{fig:P_1_4} presents the evolutionary tracks of the mass ratio, the orbital period, and the mass transfer rate for MS binary systems with initial masses $(M_1, M_2)=(3.0, 2.0)~M_\odot$. The initial orbital periods and surface magnetic fields of the evolved stars for four different evolutionary tracks are $(P_{\rm i}/{\rm days}, B_{\rm s}/{\rm G})=(0.68, 0)$, $(1.4, 2000)$, $(1.4, 5000)$, and $(1.4, 15000)$. For the three cases with non-zero magnetic fields, the anomalous MB drives continuous orbital shrinkage through efficient AM loss before Roche-lobe overflow (RLOF). A stronger magnetic field produces a higher loss rate of AM, thereby inducing an earlier RLOF: the three evolved stars with $B_{\rm s}=2000$, $5000$, and $15000~\rm G$ fill their Roche lobes at stellar ages $t=t_{\rm rlof}=215$, $164$, and $90~\rm Myr$, respectively. Since the initial mass ratio $q>1$, the mass transfer proceeds first on a short thermal timescale ($\sim1~\rm Myr$). The mass-transfer rates gradually climb and reach a peak of $\sim 10^{-6}~M_\odot\,\rm yr^{-1}$ at the first minimum period. Because the mass is transferred from the more massive star I to the less massive star II, the orbital periods gradually decrease during $q>1$. As a consequence, it should evolve to the first minimum period at $q=1$. However, our detailed binary evolution model indicates that the first minimum period exhibits a mass ratio slightly greater than 1 (see also Figure \ref{fig:P_q}). This phenomenon arises from the influence of the stellar winds of the evolved star (see also Section \ref{subsection: Wind}).

Once $q<1$, the mass transfer transitions to a long nuclear timescale ($\sim1~\rm Gyr$), and the mass-transfer rates sharply decline to $\sim 10^{-10}\mathendash 10^{-9}~M_\odot\,\rm yr^{-1}$. 
The mass transfer from the less massive star I to the more massive star II (i.e., $q<1$) would induce an orbital expansion effect. However, AM loss driven by the anomalous MB would induce an orbital decay effect. Whether Algol binaries exhibit a net orbital decay tendency critically depends on the dynamical competition between the above two effects. Because $\dot{J}_{\rm MB}\propto B_{\rm s}$, a weak magnetic field (e.g., $B_{\rm s}=2000~\rm G$) results in a low rate of AM loss. Consequently, after experiencing a platform stage (lasting hundreds of Myr), the orbital period increases rapidly during the final evolutionary stage. In contrast, the two evolutionary tracks with stronger magnetic fields of $5000$ and $15000~\rm G$ exhibit an orbital decay stage with long timescales of $\tau\sim1$ and $\sim4~\rm Gyr$, respectively. These two timescales of orbital decay are much longer than that (hundreds of Myr) given by the circumbinary (CB) disk model \citep{Chen2006}. \add{According to our simulated timescale of orbital decay, the mean orbital decay rate can be estimated as $\dot{P}\sim-P/\tau\sim-(10^{-10}-10^{-9})~\rm days\,yr^{-1}$, which is much smaller than those observed in several Algol systems (see also section 3.3).}

Considering the standard MB law, the purple curve in Figure 2 shows the evolution of an \add{MS binary} system with the orbital period (0.675 days) at the onset of RLOF, which is the same as the case with $B_{\rm s}=15000~\rm G$. The evolved star can only develop a convective envelope as $M_1\approx1.5~M_\odot$, thus the initial mass transfer rate is a bit smaller than the three cases under the anomalous MB law (this lower mass transfer rate leads to a longer mass transfer timescale, $\sim 2~\rm Myr$ when $q > 1$). Due to a low rate of AM loss, the orbital expansion effect caused by mass transfer exceeds the orbital decay effect due to AM loss, resulting in the orbital period continuously increasing when $q < 1$. Compared with the anomalous MB mechanism, the standard MB law cannot produce an orbital decay with a long timescale when $q<1$. Furthermore, MS stars without chemical peculiarities rarely sustain kilogauss-level subsurface flux tubes ($B \sim\rm kG$) capable of driving magnetic activity cycles due to insufficient convective shear and turbulent dynamo amplification. Therefore, the standard MB law is invalid for those Algol systems with orbital decay.

\begin{figure}
  \centering
  \includegraphics[width=1\columnwidth,trim={0 0 0 0},clip]{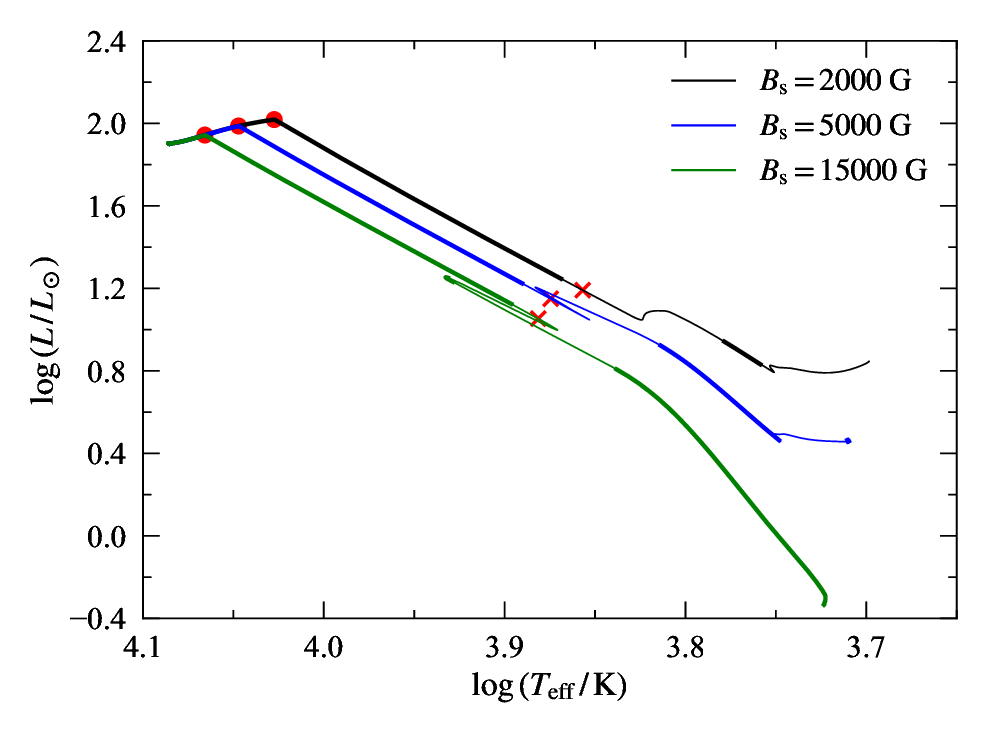}
  \caption{Same as in Figure \ref{fig:P_q}, but for the evolution of evolved stars in the HR diagram. The bold curves represent the orbital decay stages. The red solid circles and red crosses represent the onset at which the RLOF and the reversal of mass ratio take place, respectively.} \label{fig:HR_1_4}
\end{figure}

Figure \ref{fig:HR_1_4} shows the evolutionary tracks of the evolved stars in the HR diagram for the latter three binary systems presented in Figure \ref{fig:P_1_4}. Following RLOF, these three evolved stars rapidly traverse the Hertzsprung gap, then initiate to widen the orbits and achieve the reversal of mass ratios. For the same effective temperature $T_{\rm eff}$, the star with weaker surface magnetic field $B_{\rm s}$ results in higher stellar luminosity $L$. This stems from the dependence of AM loss rates on $B_{\rm s}$ via the anomalous MB: a weaker magnetic field reduces AM loss, leading to a wider orbital separation, thereby enabling the evolved star to expand to a larger radius before filling its Roche lobe. Consequently, it naturally gives rise to a higher stellar luminosity according to the Stefan-Boltzmann law ($L=4\pi R_1^2\sigma T_{\rm eff}^4$). In the second orbital decay stage, the stellar effective temperatures are ${\rm log}(T_{\rm eff}/K)=3.75\mathendash 3.84$ (i.e., $T_{\rm eff}=5600\mathendash 7000~\rm K$), which correspond to the observed spectral types F and G of the evolved stars in Algol binaries.

\begin{figure*}
  \centering
  \includegraphics[width=1\textwidth,trim={0 0 0 0},clip]{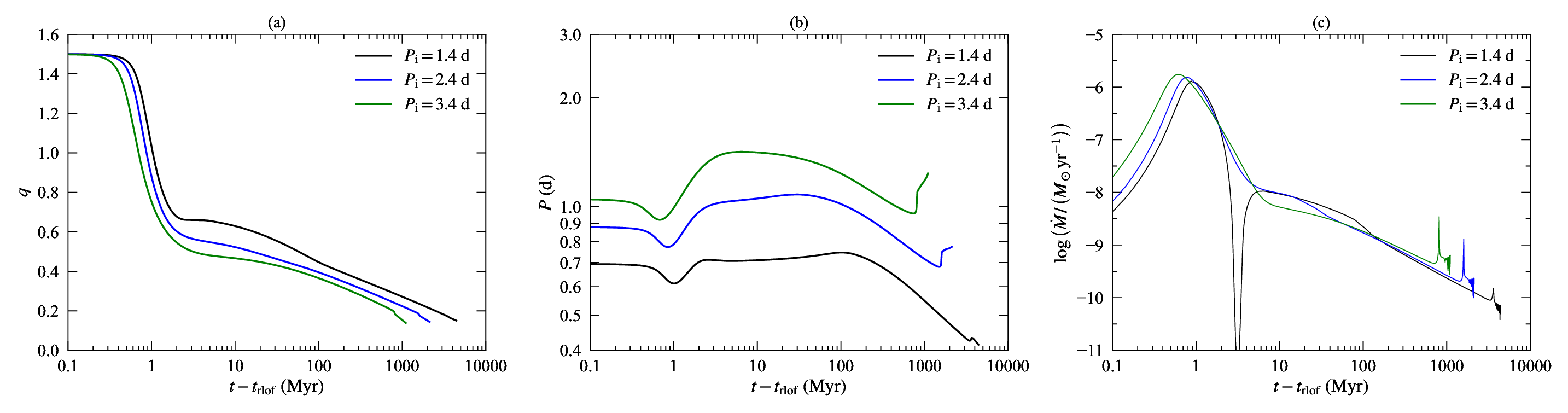}
  \caption{Same as in Figure \ref{fig:P_1_4}, but only for systems characterized by evolved stars with fixed surface magnetic field $B_{\rm s} = 13000~\rm G$. The black, blue, and green curves correspond to the initial orbital periods $P_{\rm i} = 1.4$, $2.4$, and $3.4~\rm days$, respectively.} \label{fig:B_13000}
\end{figure*}

\subsection{Influence of initial orbital periods}
Similar to Figure \ref{fig:P_1_4}, Figure \ref{fig:B_13000} illustrates the evolution of three MS binary systems characterized by evolved stars with surface magnetic fields $B_{\rm s} = 13000~\rm G$ and initial orbital periods $P_{\rm i} = 1.4$, $2.4$, and $3.4~\rm days$. The binary with a longer initial orbital period requires a longer timescale to initiate RLOF. Consequently, its donor star reaches a more advanced evolutionary state, transferring surface H-rich material to the accretor at a rate marginally higher than the two binaries with relatively shorter initial periods (see panel c). This rapid mass transfer drives the binary with a longer $P_{\rm i}$ to reach the first minimum period and the mass ratio of 1.0 sooner than the other two binaries (see panels a and b). With the onset of nuclear timescale mass transfer, the mass-transfer rates decline rapidly. Consequently, the orbital shrinkage driven by anomalous MB dominates over the orbital expansion from the mass transfer, leading to a continuous decrease in the orbital period. For binary with $P_{\rm i}=3.4~\rm days$, the mass-transfer rate at $t-t_{\rm rlof}\sim 10~\rm Myr$ is lower than that in the other two cases (see panel c), causing its orbital period to begin to decrease earlier (see panel b). At the second minimum orbital periods, the mass-transfer rates of the three systems reach their second peaks. Notably, the system with a shorter initial orbital period exhibits a longer timescale of orbital decay (see panel b). Specifically, for $P_{\rm i}=1.4$, $2.4$, and $3.4~\rm days$, the evolutionary timescales of the three systems during the second orbital decay stage are approximately $3.4$, $1.5$, and $0.7~\rm Gyr$, respectively.

\begin{figure}
  \centering
  \includegraphics[width=1\columnwidth,trim={0 0 0 0},clip]{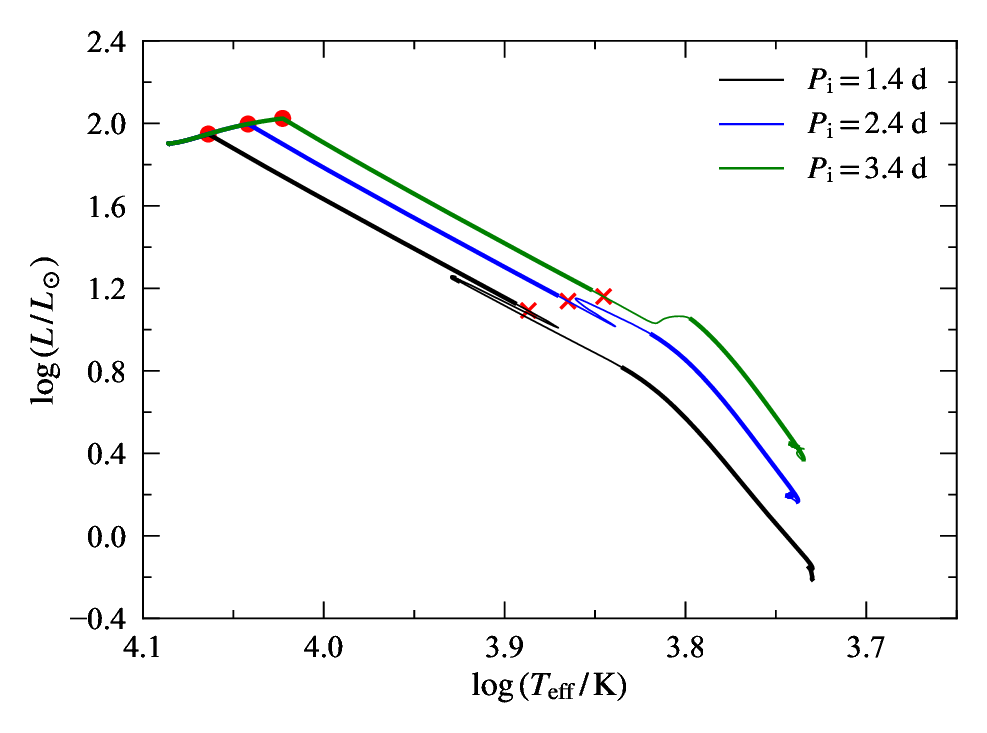}
  \caption{Same as in Figure \ref{fig:HR_1_4}, but only for systems characterized by evolved stars with fixed surface magnetic field $B_{\rm s} = 13000~\rm G$. The black, blue, and green curves correspond to initial orbital periods $P_{\rm i} = 1.4$, $2.4$, and $3.4~\rm days$, respectively. } \label{fig:HR_13000}
\end{figure}

Figure \ref{fig:HR_13000} shows the evolutionary tracks of the evolved stars in the HR diagram for the three binary systems presented in Figure \ref{fig:B_13000}. Similarly to Figure \ref{fig:HR_1_4}, the three evolved stars rapidly traverse the Hertzsprung gap during the first orbital decay stage. For the same effective temperature $T_{\rm eff}$, a longer initial period $P_{\rm i}$ results in a higher luminosity $L$ of the evolved star. This phenomenon originates from the dependence of the evolved star's radius on the initial orbital period $P_{\rm i}$ during mass transfer. A longer $P_{\rm i}$ corresponds to a larger Roche lobe, which allows the star to expand to a greater radius before RLOF, thereby increasing its luminosity $L$. In contrast, systems with shorter $P_{\rm i}$ exhibit a wider effective-temperature range in the second orbital decay stage, a result of their extended evolutionary timescales enabling prolonged thermal adjustments.

\begin{table*}
  \centering
  \caption{Some main observed parameters of five Algol-type binaries exhibiting decreasing orbital periods}
  \resizebox{\textwidth}{!}{
  \begin{threeparttable}
    \begin{tabular}{l*{8}c}
      \hline \hline
      Sources & $P$ & $\dot{P}$ & Spectral type& $M_1$ & $M_2$ & $\log L_1$ & $\log L_2$ & References\tnote* 
      \\[-0.5ex]
      & $(\rm days)$ & $(10^{-7}\rm days\,yr^{-1})$ &  & $(M_\odot)$ & $(M_\odot)$ & $(L_\odot)$ & $(L_\odot)$ &
      \\\hline
      X Tri & $0.972$ & $-1.42$ & A3/G3 & $1.2\pm 0.3$ & $2.3\pm 0.7$  & $0.45 \pm 0.18$ & $1.21\pm 0.06$ & (1)(2)\\
      AT Peg & $1.146$ & $-6.38$ & A4/G & $1.05\pm 0.025$ & $2.22\pm 0.065$ & $0.38\pm 0.07$ & $1.19\pm 0.025$ & (3)(4)\\
      AF Gem & $1.244$ & $-0.992$ & B9/G0 & $1.155\pm 0.04$ & $3.37\pm 0.11$ & $0.75 \pm 0.03$ & $1.78\pm 0.09$ & (5)(6)\\
      TX UMa & $3.063$ & $-7.13 $ & B8/G0 & $1.18\pm 0.04$ & $4.76\pm 0.16$ & $1.17 \pm 0.065$ & $2.30\pm 0.04$ & (7)(8)\\
      TW Cas & $1.428 $  & $-0.80 $ & B9/F9 & $1.15\pm 0.05$ & $2.65\pm 0.1$ & $0.98 \pm 0.069$ & $1.83\pm 0.065$ & (9)\\\hline
    \end{tabular}
    \begin{tablenotes}
      \footnotesize
      \item[*] (1) \citet{Mezzetti1980XTri}; (2) \citet{Qian2002}; (3) \citet{Maxted1994ATPeg}; (4) \citet{Qian2000a}; (5) \citet{Maxted1995AFGem}; (6) \citet{Chambliss1982AFGem}; (7) \citet{Maxted1995TXUMa}; (8) \citet{Qian2001c}; (9) \citet{Narita2001TWCas}.
    \end{tablenotes}
  \end{threeparttable}}
  \label{table:Algol_source}
\end{table*}


\subsection{A comparison between theoretical simulations and observations}
\begin{figure}
  \centering
  \includegraphics[width=1\columnwidth,trim={0 0 0 0},clip]{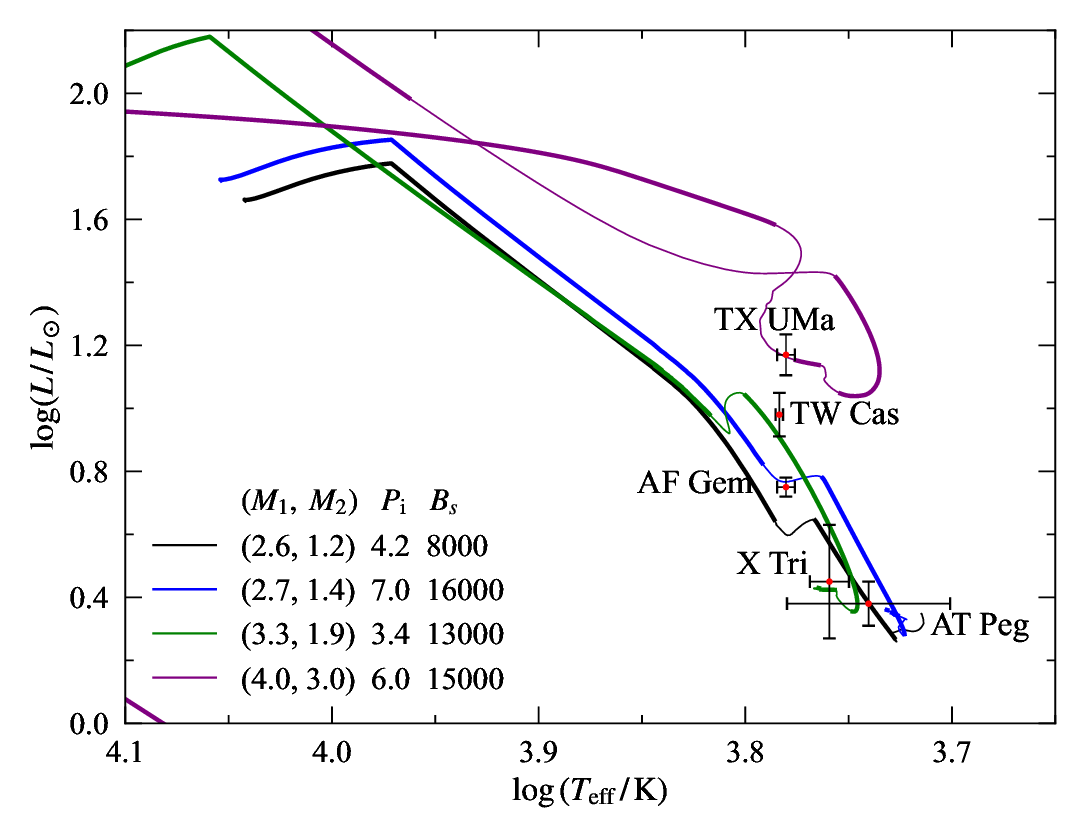}
  \caption{Evolution of the evolved stars in four MS binary systems in the HR diagram. The black, blue, green, and purple curves correspond to the evolutionary tracks of MS binary systems A$(M_1/M_\odot, M_2/M_\odot, P_{\rm i}/{\rm days}, B_{\rm s}/{\rm G}=$~2.6, 1.2, 4.2, 8000), B(2.7, 1.4, 7.0, 16000), C(3.3, 1.9, 3.4, 13000), and D(4.0, 3.0, 6.0, 15000), respectively. The bold curves represent the orbital decay stages. The solid red circles with error bars denote the locations of the five Algol binaries in Table \ref{table:Algol_source}.} \label{fig:HRs}
\end{figure}

In order to test the anomalous MB model, we compare our simulated results with the observations. Table \ref{table:Algol_source} summarizes some typical parameters of five Algol systems with orbital decay. Given the observed sign reversal in TW Cas's orbital period derivative $\dot{P}$ (\add{\citealp{Ma2022TWCas} found that $\dot{P}$ of TW Cas is no longer negative through the $O-C$ curve analysis, but $1.41\times 10^{-7}~\rm days\,yr^{-1}$}), 
we omitted detailed fitting in the comparison diagram and instead marked it as a comparison with the other four sources. Based on the effective temperature range which was derived according to the stellar spectral types \citep{Pecaut2013}, Figure \ref{fig:HRs} gives a comparison between our simulated results and the observations in the HR diagram for four different initial parameters. In the orbital decay stage, the evolutionary curves of the systems A, B, and D can match the observed parameters of X Tri, AT Peg, and TX UMa, respectively. For AF Gem, the radius of the evolved star calculated by the system C is slightly larger than the observed value (i.e., for the same effective temperature $T_{\rm eff}$, system C yields a higher luminosity $L$ than observed). For the CB disk model, the simulated results show good agreement with the observations of four Algol systems in the HR diagram \citep{Chen2006}.



\begin{figure}
  \centering
  \includegraphics[width=1\columnwidth,trim={0 0 0 0},clip]{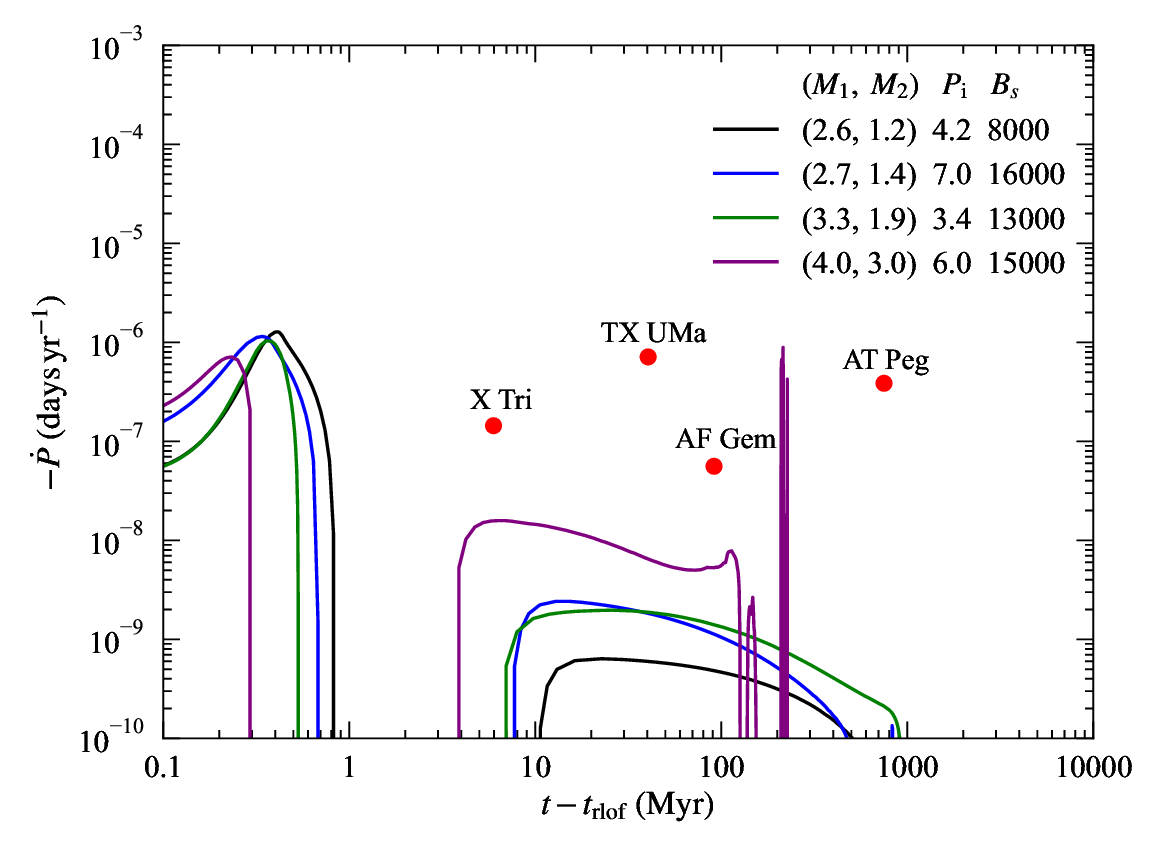}
  \caption{Same as in Figure \ref{fig:HRs}, but for a comparison between our simulated results and the observed parameters in $-\dot{P}$ vs. the mass-transfer timescale ($t-t_{\rm rlof}$) diagram. } \label{fig:Pdot_t}
\end{figure}

\begin{figure}
  \centering
  \includegraphics[width=1\columnwidth,trim={0 0 0 0},clip]{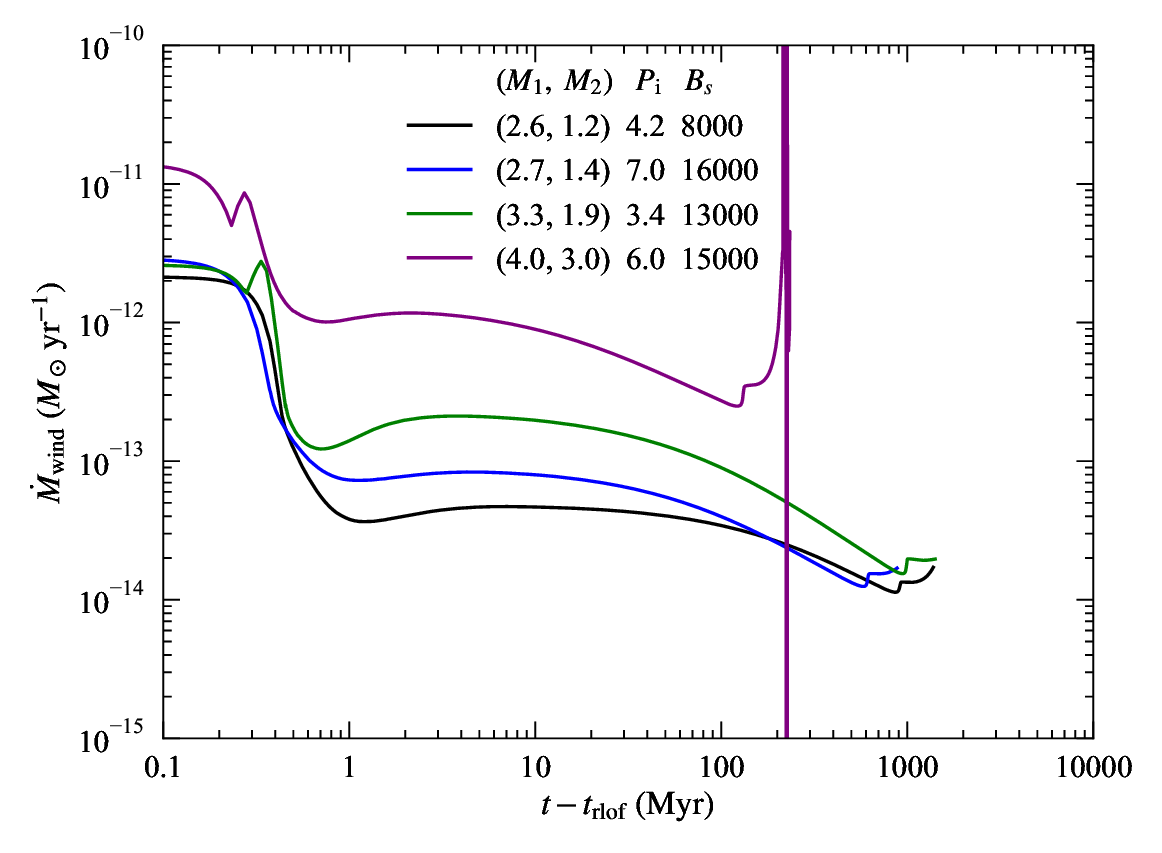}
  \caption{Same as in Figure \ref{fig:HRs}, but for the evolution of the stellar wind loss rates of evolved stars as a function of the mass-transfer timescale ($t-t_{\rm rlof}$). } \label{fig:Wind}
\end{figure}

Figure \ref{fig:Pdot_t} depicts a comparison between the predicted results and the observations in $-\dot{P}$ versus the mass-transfer timescale diagram. For simplicity, we only show the evolutionary tracks in the orbital decay stage. When $q > 1$, the simulated orbital decay rates are as high as $\sim 10^{-7}\mathendash 10^{-6}~\rm days\,yr^{-1}$. Once $q<1$, the period derivatives $-\dot{P}$ decrease significantly to $\sim 10^{-10}\mathendash 10^{-8}~\rm days\,yr^{-1}$, which are $2\mathendash 3$ orders of magnitude smaller than the values observed in these sources. The sharp decline of the orbital decay rates originates from a fast decay of stellar wind loss rates (see also Figure \ref{fig:Wind}). Therefore, the anomalous MB mechanism could potentially reproduce the basic parameters of Algol systems but could not contribute to the rapid orbital decay rates observed in these systems. For the CB disk model, \cite{Chen2006} did not perform a comparison between theoretically predicted period derivatives and the observations.  Recently, \cite{fan24} found that the convection- and rotation-boosted MB law can produce a maximum negative period derivative of $-\dot{P}\approx10^{-10}~\rm s\,s^{-1}$ in neutron star (or black hole) low-mass X-ray binaries. However, such a period derivative of $-\dot{P}\approx4.0\times10^{-8}~\rm days\,yr^{-1}$ is still too small to account for the orbital decay rates of Algol systems.

\section{Discussion} \label{section:Discussion}
\subsection{Limitations of the anomalous MB model}
\add{Based on the anomalous MB model of Ap/Bp stars proposed by \citet{Justham2006}, our detailed binary evolution calculations indicate that the MB mechanism of strongly magnetic ($\sim10^4~\rm G$) Ap/Bp stars can efficiently extract orbital angular momentum and lead to orbital decay, however the decay timescales are from hundreds of Myr to several Gyr, which is 2-3 orders of magnitude longer than orbital decay timescales ($\tau_{\rm decay}\sim -P/\dot{P}\sim10~\rm Myr$) inferred from the observed period $P$ and period derivative $\dot{P}$ of several Algol binaries (assuming that the observed period changes represent a secular trend). This is also confirmed by comparisons between the predicted results and the observations in $-\dot{P}$ versus the mass-transfer timescale diagram (see Figure \ref{fig:Pdot_t}). As for why MS binaries containing Ap/Bp stars are unlikely to evolve into Algol systems with rapid orbital decay, the sharp decline in stellar wind loss rates should be a key factor (see Figure \ref{fig:Wind}). According to Equation (3), a lower wind-loss rate produces a lower rate of angular momentum loss through the anomalous MB mechanism, naturally resulting in a longer evolutionary timescale. Furthermore, some known characteristics of Ap/Bp stars may be unfavorable for the anomalous MB model, as well. 
\begin{enumerate}
    \item Ap/Bp stars in close spectroscopic binaries are found to be rare. In a survey of 151 close ($P_{\rm orb} < 20~\rm d$) intermediate-mass or massive binary systems with magnitude $V < 8$ mag, \citet{Alecian2015BinaMIcS} detected only one magnetic field, in the secondary component of HD~160922. In particular, Ap stars are very rare in short-period binary systems with a mass ratio similar to that of Algol systems \citep{fols13}. This rarity makes it highly unlikely that Algol systems, which are interacting binaries, originate from such configurations.
    \item Ap/Bp stars are generally slow rotators. Based on the fact that not all slowly rotating A2 to B5 stars exhibit chemical peculiarities, \citet{taur23} suggested that the magnetic fields of Ap/Bp stars may decay on timescales of $\sim 10^8~\rm yr$.
    Even neglecting magnetic field decay, our binary evolution calculations for MS binaries containing Ap/Bp stars cannot reproduce the rapid orbital decay rates observed in those Algol systems. If the magnetic field decay is taken into account, the disparity between the predicted and detected orbital period derivatives will be further widened.
\end{enumerate}}
The existence of Ap/Bp stars may be confirmed through the observation of spectral line splitting resulting from the Zeeman effect. For those Ap/Bp stars with low projected rotating velocities and strong magnetic fields, the wavelength separations of resolved, magnetically split lines (RMSLs) can be used to measure the mean magnetic field modulus. Recently, magnetic field measurements of 84 Ap stars and 157 Ap/Bp stars with RMSLs were reported by \cite{math17} and \cite{Chojnowski2019}, respectively. In more than half of Ap stars with RMSLs, the measured magnetic field strengths lie in the range of $(4\mathendash 7)\times 10^3~\rm G$ \citep{Chojnowski2019}. We anticipate that ongoing large-scale surveys \textemdash{} specifically the Sloan Digital Sky Survey (SDSS) and Apache Point Observatory Galactic Evolution Experiment (APOGEE) \textemdash{} \add{will further increase the sample size of Ap/Bp stars, which would help confirm their rarity in close spectroscopic binaries and constrain the timescale of their magnetic field decay.}

\subsection{Other potential mechanisms}
\subsubsection{A surrounding CB disk}
The tidal torque between an Algol system and a surrounding CB disk can efficiently extract the orbital AM from the binary, causing the system to experience an orbital decay on a sufficiently long timescale \citep{Chen2006}. However, their work had not calculated the orbital period derivatives of Algol systems. \add{There exist a rapid orbital decay stage with a timescale of $\tau\sim10~\rm Myr$ when the mass transfer time $t= 200-300~\rm Myr$ in Figure 2 of \cite{Chen2006}. An orbital decay rate can be approximately estimated as $\dot{P}\sim P/\tau\sim10^{-7}~\rm days\,yr^{-1}$, implying that a surrounding CB disk could  produce an orbital period derivative matching the observations. } According to the torque exerted on the binary, the thermal luminosity of a CB disk can be estimated to be $L_{\rm cb}=-2\pi\dot{J}_{\rm cb}/P$ \citep{spru01}. Equation (\ref{eq:pdot}) yields $-\dot{J}_{\rm cb}/J>-\dot{P}/3P$ because mass transfer produces a positive $\dot{P}$ in an Algol binary. Therefore, we have $L_{\rm cb}>-2\pi\dot{P}J/(3P^{2})$. For X Tri, the minimum luminosity can be derived as $L_{\rm cb}=2.5\times10^{32}~\rm erg\,s^{-1}$. Since the CB disk is mainly radiated in a zone near the inner edge \citep{spru01}, the effective emitting area can be estimated to be $A\sim\pi\gamma^2a^2=1.3\times10^{24}~\rm cm^2$, in which we take $\gamma=1.5$ \citep{sobr97}. According to the Stefan–Boltzmann law, the minimum effective temperature at the inner edge of the CB disk is derived to be $T_{\rm eff}=1360~\rm K$, implying that the CB disk could be detected in the near-infrared band if it exists \citep{gaia20}. 

\subsubsection{Shell expansion of the evolved star}
The shell expansion of a late-type evolved star could cause an increase in the moment of inertia, leading to a spin-down of the stellar rotation. The tidal interaction between the two components would spin the evolved star back up to corotate with the orbital motion, indirectly extracting orbital AM from the binary. This star expansion model could be responsible for the orbital decay observed in the high-mass X-ray binary LMC X-4 \citep{levi00}. Unfortunately, this model is invalid for Algol systems with mass transfer via RLOF. Because the radius of the evolved star is equal to its Roche-lobe radius, it continuously decreases due to orbital decay, resulting in a shrinkage rather than an expansion of the evolved star.  

\subsubsection{Magnetic activity cycles}
\add{If the observed orbital decay in several Algol systems is merely a short-term phenomenon, the Applegate mechanism \textemdash{} which can induce modulations of the orbital period on timescales of decades or longer \textemdash{} may provide an alternative explanation.} The shear caused by the differential rotation and the Coriolis force drives a magnetic activity cycle in a poloidal and toroidal field. This magnetic activity transfers angular momentum in different convective zones of the star to cause a variable deformation \citep{appl92}. The change in the moment of inertia induces a variation of the spin period of the star, and hence the orbital period of the Algol system due to the tidal locking. For orbital period modulations of amplitude $\Delta P/P\sim 10^{-5}$, it requires a mean subsurface magnetic field of $\sim10^{3}~\rm G$ for the torque to redistribute the angular momentum \citep{appl92}. \add{\citet{Yang2014AFGem} proposed that the orbital period of AF Gem may exhibit a cyclic variation, potentially driven by magnetic activity cycles of the donor star with a change ($\Delta Q=1.54\left(\pm 0.06\right)\times 10^{50}~\rm g\,cm^2$) of the gravitational quadrupole moment. Furthermore, the period derivative $\dot{P}$ of TW Cas transitioned from negative values \citep[$-0.8\times10^{-7}~\rm days\,yr^{-1}$,][]{Narita2001TWCas} to positive values \citep[$1.41\times10^{-7}~\rm days\,yr^{-1}$,][]{Ma2022TWCas}, also providing evidence for the cyclic variation of orbital period. However, \citet{Ma2022TWCas} only attempted to explain the cyclic variation superimposed on the long-term orbital period increase of TW Cas using the Applegate mechanism, and this explanation failed
because the total energy radiated by the donor star in the oscillating period of 113.41 years was insufficient to match the required energy of this mechanism. As for the cause of the sign reversal in $\dot{P}$ of TW Cas, it remains uncertain and requires further investigation.}

\add{Moreover, it is worth noting that the Applegate mechanism relies on magnetic activity cycles driven by dynamo effect, which involve angular momentum redistribution within the stellar convective envelope. However, \citet{Braithwaite2017} proposed that stable Ap-type magnetic fields cannot be sustained in stars with convective envelopes, and \citet{Hidalgo2024} suggested that the magnetic fields generated by the convective core dynamo are insufficient to account for the observed surface magnetism in Ap/Bp stars. Consequently, the Applegate mechanism of Ap/Bp stars is not applicable as an alternative explanation for the orbital decay observed in several Algol systems \textemdash{} even if the observed decay represents merely a short-term phenomenon.}

\subsubsection{Light travel-time effect caused by a third body}
\add{Unlike the quasi-periodic orbital modulations produced by the Applegate mechanism, strictly periodic variations with more than one cycle are typically attributed to the presence of a third body, with the variations arising from the light-travel time effect induced by that body.
For AF Gem, \citet{Yang2014AFGem} also proposed the presence of a third body as an alternative explanation for the observed cyclic variation of the orbital period.
For TW Cas, \citet{khal15} and \citet{pari18} reported a cyclic variation of the orbital period based on the $O-C$ diagram without any long-term period change, and attributed it to the presence of a third body successfully. Recently, \citet{Ma2022TWCas} analyzed the $O-C$ diagram of TW Cas based on light curves and 64 eclipsing times observed by Transiting Exoplanet Survey Satellite \citep[TESS,][]{Ricker2015TESS} telescope, and found that a cyclic variation is superimposed on a long-term increase of the orbital period.
They successfully explained the long-term increase through the rapid mass-transfer process and the superimposed cyclic variation through a third body.
The strictly periodic variations observed in the orbital periods of Algol systems are usually well explained by invoking the presence of a third body. Photometric studies and model fittings suggest that the third body typically has a long orbital period or is a cold, faint stellar object, contributing negligibly to the total luminosity of a system. To test the validity of this model, further high-resolution spectroscopic observations are essential. Moreover, it remains unclear whether the observed rapid orbital decay in several Algol systems is indeed merely a segment of a cyclic variation induced by a third body. Long-term observations are required in the future, as well as detailed theoretical calculations of orbital evolution involving a third body.}

\section{Summary}\label{section:Summary}
In this work, we investigate whether the anomalous MB mechanism could produce Algol systems with orbital decay. If the evolved stars evolve from those Ap/Bp stars with anomalously strong surface magnetic fields, the coupling between the magnetic field and stellar wind can cause the rotation of the evolved star to slow down, resulting in an efficient loss mechanism of orbital AM. For some typical initial parameters and a relatively strong magnetic field ($B_{\rm s}\sim10000~\rm G$), our detailed binary evolution models revealed that the orbits of MS binaries can continuously shrink over a long timescale (hundreds of Myr to several Gyr) \add{, which is nevertheless 2-3 orders of magnitude longer than inferred orbital decay timescales of those Algol systems. Our simulated results could match the observed parameters of three Algol systems in the HR diagram (see Figure \ref{fig:HRs}). However, the predicted period decay rates ($\sim 10^{-10}\mathendash 10^{-8}~\rm days\,yr^{-1}$) are much smaller than the observed values ($\sim 10^{-7}~\rm days\,yr^{-1}$, see Table \ref{table:Algol_source}). These results suggest that Ap/Bp stars in MS binaries could, in principle, evolve into donor-like stars similar to those found in Algol systems. However, the anomalous MB mechanism cannot produce the orbital decay rates observed in several Algol systems. 
Furthermore, both the rarity of Ap/Bp stars in close spectroscopic binaries and the surface magnetic field decay timescale of $\sim 10^8~\rm yr$ imply that the anomalous MB mechanism cannot be responsible for the rapid orbital decay observed in several Algol systems, as well.}

\add{It seems that the observed long- or short-term orbital decay in several Algol systems can be interpreted by three mechanisms, including a surrounding CB disk, magnetic activity cycles of evolved stars, and a third body. For the CB disk model, near-infrared spectroscopic observations are usually required to verify their existence. Regarding the magnetic activity cycles model, it is important to note that the Ap/Bp stars studied in this work \textemdash{} characterized by intermediate-mass and radiative envelopes \textemdash{} are not applicable to this mechanism. As for a third-body model, the additional tertiary companion \textemdash{} usually characterized by either a long orbital period or very low temperature and luminosity \textemdash{} requires confirmation through high-resolution spectroscopic observations. Moreover, the period changes caused by the magnetic activity cycles or by a third body are periodic rather than secular. Therefore, these two mechanisms can potentially be distinguished from genuine mass/AM loss, such as that caused by a CB disk.
Long-term further observations are required to confirm or rule out these mechanisms in the future.}

\acknowledgments{We are incredibly grateful to the referee for
constructive comments that improved this manuscript. This work was partly supported by the National Natural Science Foundation of China (under grant Nos. 12273014 and 12473043) and the Natural Science Foundation (under grant number ZR2021MA013) of Shandong Province.}

\bibliography{reference}{}
\bibliographystyle{aasjournal}

\end{document}